\documentstyle[12pt]{article}
\begin{document}
\thispagestyle{empty}
\vskip 2cm
\begin{flushright}
KUCP/U-0081 \\
TIT/HEP-296/COSMO-55
July 1995
\end{flushright}
\vskip 1cm
\begin{center}
{ \huge  Hamilton-Jacobi Equation for Brans-Dicke Theory and
Its Long-wavelength Solution} \\
\vskip 2cm
          { \large \bf  Jiro Soda }

\vskip 0.5cm
      Department of Fundamental Sciences \\
          FIHS, Kyoto University, Kyoto 606, Japan \\
\vskip 1cm
      {\large \bf Hideki Ishihara and Osamu Iguchi } \\
      Department of Physics \\
      Tokyo Institute of Technology, Meguroku, Tokyo 152, Japan
\end{center}

\vskip 5cm

\begin{abstract}
 Hamilton-Jacobi equation for Brans-Dicke theory is solved by using a
long-wavelength
 approximation. We examine the non-linear evolution of the inhomogeneities in
the dust
 fluid case and the cosmological constant case.
 In the case of dust fluid, it turns out that the inhomogeneities of space-time
grow.
 In the case of cosmological constant, the inhomogeneities decay, which is
consistent with the cosmic no hair conjecture. The inhomogeneities of the
density perturbation and
 the gravitational constant behave similarly with that of space-time.
\end{abstract}

\newpage

\baselineskip 0.8cm

\section{Introduction}
It is generally believed that the consistent theory of quantum gravity is
correctly described by
the superstring theory. The classical theory of gravity is nothing but a low
energy effective theory of the
 superstring theory.  The classical theory which is predicted by the
superstring theory has the form
 of a scalar-tensor theory, the scalar field is the so-called dilaton.
Therefore, it is necessary
 to consider the consequences of this extra scalar field at least in the
phenomena close to the
 Planck scale. If the dilaton field acquires a large mass due to unknown
dynamical mechanism,
 there will be no observable macrospic difference between the superstring
predicted theory and
 Einstein's  general theory of relativity. However, recently, it is pointed out
the possiblity
 of the massless dilaton.\cite{damour}
 If so, it is important to study scalar-tensor theory more seriously.
 The simplest scalar-tensor theory is Brans-Dicke theory\cite{BD}
 where the dilaton field acts like a
 dynamical gravitational constant.

 On the other hand, to circumvent the graceful-exit problem of old inflation,
Brans-Dicke theory
 is renewed in the inflational universe scenario as extend inflation.
Moreover, Bellido et.al.\cite{linde}
 investigate the stochastic inflation formalism in the context of Brans-Dicke
theory.
 They described the inhomogeneous universe with fluctuations of the
gravitational constant.
 The main idea of stochastic inflation is to solve the equations for the
inhomogeneous fields
 in de-Sitter space by separating both the gravitational and scalar fields in
short wavelength
 quantum fluctuations, which oscillate on scales smaller than the Hubble
radius, and long wavelength fluctuations which are treated as classical fields.
Salopek and Bond\cite{SB} have developed the
 formalism to treat the long wavelength fluctuations in terms of
Hamilton-Jacobi equation,
 which is applicable not only to the inflational theory but also the late stage
evolution
 of the density fluctuations as far as the typical scale of the fluctuation
exceeds the Hubble
 radius.

  The so-called long wavelength approximation has rather long history dating
back to
Lifshitz and Khalatonikov.\cite{LK} Later,
Tomita developed the above approximation as the Anti-Newtonian
 scheme.\cite{tomita}
Recently, Salopek and co-orthors elegantly formulated the long wavelength
approximation
 in the context of Hamilton-Jacobi theory.\cite{PSS,CPSS}
 The direct method of Comer et. al.\cite{comer} is also useful
 to calculate the higher order correction.
 Attempts to apply the formalism to the inflationary theory and the higher
dimensional
 theory also exist.\cite{N}

 In this paper, we shall apply the long wavelength approximation to Brans-Dicke
theory.
We are interested in the non-linear evolution of the long-wavelength
inhomogeneities.
 As the inhomogeneties of the Brans-Dicke field implies the inhomogeneities
 of the gravitational constant, it is important for the astrophysical
phenomena.
 In our modest study, of course, we do not intend to make definite statements
 to the astrophysics. However, it is important to investigate how the
inhomogeneities
 of the gravitational coupling constant evolve.
 We start in Sec.2 by writing the Hamilton-Jacobi equation for Brans-Dicke
theory.
The spatial gradient expansion is performed in Sec.3.
 As a matter field, the cosmological constant is interesting. This case is
related
 to the cosmological no-hair conjecture. Sec.4 is devoted to these subjects.
 In Sec.5, the discussions of the various problems are presented. In the
Appendix,
 the results of the direct method are explained.

\section{Hamilton-Jacobi Equation for Brans-Dicke Theory}

For simplicity, we will consider the dust fluid matter.
Action for the Brans-Dicke theory with the dust matter $\chi$
is given by
\begin{equation}
I= \int d^4x \sqrt{-g} [ \phi ^{(4)}R - {\omega \over \phi} g^{\mu\nu}
     \partial_\mu \phi  \partial_\nu \phi
      - {n \over 2m} ( g^{\mu\nu} \partial_\mu \chi \partial_\nu \chi + m^2) ]
\ ,
\end{equation}
where $n$ is the Lagrange multiplier and $m$ is the particle mass which is
normalized
 to unity below. Here, $\omega$ is the parameter of the theory.
 The Brans-Dicke field $\phi$ is considered as the effective
 gravitational coupling constant.
The Hamilton-Jacobi equation for Brans-Dicke theory is obtained using the
 Arnowitt-Deser-Misner (ADM) formalism in which the space-time is foliated by
 a family of space-like hypersurfaces.
In the ADM formalism, the metric is parametrized as
\begin{equation}
ds^2 = -N^2 dt^2 + \gamma_{ij} (dx^i + N^i dt)(dx^j + N^j dt)  \ ,
\end{equation}
where $N$ and $N^i$ are the lapse and shift functions respectively, and
$\gamma_{ij}$ is the 3-metric.
Using the above metric, we obtain the Hamiltonian form of the action as
\begin{equation}
I=\int d^4x \{ \pi^{ij} \dot\gamma_{ij} +\pi^\phi \dot\phi + \pi^\chi \dot\chi
             - N {\it H} -N^i {\it H_i}   \}  \ ,
\end{equation}
where
\begin{eqnarray}
  {\it H} &=& {1\over \phi \sqrt{\gamma} } \pi^{ij} \pi^{kl}
          [\gamma_{ik} \gamma_{jl} - {\omega +1 \over 2\omega +3} \gamma_{ij}
\gamma_{kl}]
          + {1\over 2(2\omega +3) } {\phi \over \sqrt{\gamma}} \pi_\phi^2
          - {1\over 2\omega +3} {1\over \sqrt{\gamma} } \pi \pi_\phi  \nonumber
\\
        & &   + \sqrt{1+\gamma^{ij} \chi_{,i} \chi_{,j} }  \ \pi^\chi -
\sqrt{\gamma} \phi R
           + \omega {\sqrt{\gamma} \over \phi} \gamma^{ij}
               \partial_i \phi \partial_j \phi
           + 2\sqrt{\gamma} \Delta \phi  \\
    {\it H_i }     &= & -2 (\gamma_{ik} \pi^{kj})_{,j} + \pi^{lk} \gamma_{lk,i}
                   \pi^\phi \phi_{,i} + \pi^\chi \chi_{,i}  \ .
\end{eqnarray}
Here $\pi^{ij}, \pi^{\phi}$ and $\pi^{\chi}$ are conjugate to $\gamma_{ij},
\phi$ and $\chi$, respectively,
and $R$ denotes the 3-dimensional scalar curvature.
Variation with respect to the momentum yields the equations of motion
\begin{eqnarray}
 {1\over N} (\dot \phi - N^i \phi_{,i} ) &=& {1\over 2\omega +3} {1\over
\sqrt{\gamma}}
                 [\phi \pi_\phi - \pi ] \ , \\
{1\over N} (\dot \chi - N^i \chi_{,i} ) &=&\sqrt{1+\gamma^{ij} \chi_{,i}
\chi_{,j} } \ , \\
{1\over N} (\dot \gamma_{ij} - N_{i\mid j} -N_{j\mid i} ) &=&
    {2 \over \phi \sqrt{\gamma} }  \pi^{kl}
          [\gamma_{ik} \gamma_{jl} - {\omega +1 \over 2\omega +3} \gamma_{ij}
\gamma_{kl}]
         - {1\over 2\omega +3} {1\over \sqrt{\gamma} } \gamma_{ij} \pi_\phi  \
{}.
\end{eqnarray}
Variation of the action (3) with respect to the field variables yield the
evolution
 equations for the momentum. They are automatically satisfied provided that
\begin{equation}
 \pi^{ij} = {\delta S \over \delta \gamma_{ij}} \ ,  \
  \pi^\phi = {\delta S \over \delta \phi} \ , \
  \pi^\chi = {\delta S \over \delta \chi}
\end{equation}
satisfy the constraint equations and provided the evolution equations (6) hold.
Here, instead of solving the equations of motion for the momentum fields,
 we will use the Hamilton-Jacobi method.
Hamilton-Jacobi equation is
\begin{eqnarray}
 & &  {1\over \phi \sqrt{\gamma} } {\delta S \over \delta \gamma_{ij}}
           {\delta S \over \delta \gamma_{kl}}
         [\gamma_{ik} \gamma_{jl} - {\omega +1 \over 2\omega +3}
                          \gamma_{ij} \gamma_{kl}]  \nonumber \\
     & & \qquad   + {1\over 2(2\omega +3) }
           {\phi \over \sqrt{\gamma}} ({\delta S \over \delta \phi})^2
          - {1\over 2\omega +3} {1\over \sqrt{\gamma} }
             \gamma_{ij} {\delta S \over \delta \gamma_{ij} }
               {\delta S \over \delta \phi}   \\
        & & \qquad  + \sqrt{1+\gamma^{ij} \chi_i \chi_j } {\delta S \over
\delta \chi}
                - \sqrt{\gamma} \phi R
           + \omega {\sqrt{\gamma} \over \phi} \gamma^{ij} \partial_i \phi
\partial_j \phi
           + 2\sqrt{\gamma} \Delta \phi =0  \ .  \nonumber
\end{eqnarray}
The momentum constraint is rather trivial condition which states that the
generating
functional is invariant under the spatial coordinate transformation.

 Hamilton-Jacobi formalism has a great advantage, that is, its intimate
relation to the
 quantum gravity. Wheeler-DeWitt equation for Brans-Dicke theory is given by
\begin{equation}
    H \Psi =0 \ \ , \ \ \ \ H_{i} \Psi =0 \ ,
\end{equation}
where the canonical commutation relations
\begin{eqnarray}
  \left[ \gamma_{ij} (x) \ , \pi^{kl} (y)  \right] &=&
         {i \over 2}(\delta^k_i \delta^l_j +\delta^k_j \delta^l_i ) \delta(x-y)
\ , \\
 \left[ \phi(x) \ , \pi^{\phi}(y) \right] &=& i \delta (x-y) \ , \\
  \left[\chi(x) \ , \pi^{\chi}(y) \right] &=& i \delta (x-y)
\end{eqnarray}
are assumed. If we consider the WKB approximation, we get eq.(10) as the lowest
equation.
 The research in this direction from the point of view of the long-wavelength
 approximation is the issue to which we shall attack in the future.

\section{Long Wavelength Solution}

The heart of the long wavelength approximation is the following:
For illustration, we take metric in the synchronous form
\begin{equation}
  ds^2 = - dt^2 + \gamma_{ij} (x^k , t) dx^i dx^j \ \ \ (i,j=1,2,3) .
\end{equation}
At each point one  can define a local scale factor $a$ and a local Hubble
parameter $H$ by
\begin{equation}
 a^2 \equiv (\det \gamma_{ij} )^{1 \over 3}  \ \  ,   H \equiv {\dot a \over a
} \ ,
\end{equation}
where dot denotes the time derivative.
The Hubble parameter leads to the characteristic proper time on which the
metric evolves.
 The characteristic comoving length on which it varies is denoted $L :
\partial_i \gamma_{jk}
 \approx L^{-1} \gamma_{ij} $ . The long wavelength approximation is the
assumption
that the characteristic scale of spatial variation is much bigger than the
Hubble radius, that is:
\begin{equation}
  {1\over a} \partial_i \gamma_{jk} \ll \dot \gamma_{ij}
        \Longleftrightarrow  aL \gg H^{-1}  \ .
\end{equation}
Then we can drop the spatial curvature term in the Einstein equations
 in the lowest order. If we incorporate the curvature effect perturbatively,
 we get the understanding of the non-linear evolution of the inhomogeneities.
 This direct method is technically useful. We presented the analysis in the
Appendix
 for the purpose of the check. For the conceptual reason, here, we take another
 approach, i.e. , we consider the long-wavelength approximation in
 the context of the Hamilton-Jacobi formalism.

Let us follow the method developed by Salopek and the co-orthors.
 They expand the generating functional in a series of terms according to the
number
of spatial gradients they contain:
\begin{equation}
  S = S^{(0)} + S^{(2)} + S^{(4)} + S^{(6)} +   \cdots .
\end{equation}
As a result the Hamilton-Jacobi equation can be solved perturbatively as we
will show.
The lowest order Hamilton-Jacobi equation is
\begin{eqnarray}
    & & {1\over \phi \sqrt{\gamma} } {\delta S^{(0)} \over \delta \gamma_{ij}}
           {\delta S^{(0)} \over \delta \gamma_{kl}}
          [\gamma_{ik} \gamma_{jl} - {\omega +1 \over 2\omega +3} \gamma_{ij}
\gamma_{kl}]
       + {1\over 2(2\omega +3) } {\phi \over \sqrt{\gamma}} ({\delta S^{(0)}
\over \delta \phi})^2  \nonumber \\
          & & \quad  - {1\over 2\omega +3} {1\over \sqrt{\gamma} }
             \gamma_{ij} {\delta S^{(0)} \over \delta \gamma_{ij} }
               {\delta S^{(0)} \over \delta \phi}
   + \sqrt{1+\gamma^{ij} \chi_i \chi_j } {\delta S^{(0)} \over \delta \chi} =0
\ .
\end{eqnarray}
It is difficult to obtain the complete solutions of the Hamilton-Jacobi
equation.
 Fortunately, we  need only the growing mode in the late stage or inflationary
stage.
Hence, we put the qusi-isotropic ansatz
\begin{equation}
  S^{(0)} = - 2 \int d^3x \sqrt{\gamma} \phi H(\chi) \ .
\end{equation}
The diffeomorphyism invariance is automatically satisfied in this ansatz.
Substituting the ansatz (20) into eq.(19), we obtain
\begin{equation}
   H^2 = - {4\omega +6 \over 3\omega +4 } {\partial H \over \partial \chi} \ .
\end{equation}
The solution of eq.(21) becomes
\begin{equation}
 H(\chi) = {4\omega +6 \over 3\omega +4} {1\over \chi} ,
\end{equation}
where we ignored the integration constant, because it is a decaying mode.
Hereafter, we choose the comoving gauge $\chi_{,i}=0$ .
 In principle, we are free to choose the time slicing. For a dust fluid,
 a natural choice is a comoving gauge $\chi_{,i} =0 $. Fortunately, in this
 case, we can also take the synchronous gauge, and $\chi$ has a meaning of
time.
Equations of motion give
\begin{eqnarray}
 \dot \phi &=& {H \over 2\omega +3 } \phi  \ , \\
 \dot \gamma_{ij} &=& {2\omega +2 \over 2\omega +3} H \gamma_{ij} \  .
\end{eqnarray}
Their solutions are
\begin{eqnarray}
 \phi &=& \chi^{2\over 3\omega +4} \tilde \phi(x) \ , \\
 \gamma_{ij} &=& \chi^{4\omega +4 \over 3\omega +4} h_{ij}(x) \  ,
\end{eqnarray}
where $\tilde \phi(x)$ and $h_{ij} (x)$ are arbitrary functions of spatial
coordinates,
 which we call seed scalar
 and seed metric, respectively.

The second order Hamilton-Jacobi equation is given by
\begin{eqnarray}
   & & {2\omega +2 \over 2\omega +3} \gamma_{ij}{\delta S^{(2)} \over \delta
\gamma_{ij} }            + { H \over 2\omega +3 } \phi  {\delta S^{(2)} \over
\delta \phi}
                                                      \nonumber  \\
        & & \quad   +  {\delta S^{(2)} \over \delta \chi}
                - \sqrt{\gamma} \phi R
         + \omega {\sqrt{\gamma} \over \phi}
          \gamma^{ij} \partial_i \phi \partial_j \phi
           + 2\sqrt{\gamma} \Delta \phi =0  \ .
\end{eqnarray}
To simplify the Hamilton-Jacobi equation, we utilize a conformal transformation
 of the three-metric and the Brans-Dicke scalar to define variables
\begin{eqnarray}
   f_{ij}&=& \Omega^{-2} \gamma_{ij} \ ,  \\
   \psi &=&   W^{-1} \phi \ ,
\end{eqnarray}
where $\Omega$ and $W$ satisfy
\begin{eqnarray}
 { \partial \Omega \over \partial \chi } &=& {\omega +1 \over 2\omega +3} H
\Omega \ , \\
 {\partial W \over \partial \chi } &=& {1\over 2\omega +3} H W \ .
\end{eqnarray}
And then,
\begin{eqnarray}
   \Omega &=& \chi^{2\omega +2 \over 3\omega +4} \ , \\
   W &=& \chi^{2 \over 3\omega +4} \ .
\end{eqnarray}
Hence, the Hamilton-Jacobi equation reduces to
\begin{equation}
 {\delta S^{(2)} \over \delta \chi } \mid_{f_{ij},\psi} =
  - W\Omega \left[ -\sqrt{f} \psi R(f) + \omega {\sqrt{f} \over \psi}
   f^{ij} \partial_i \psi \partial_j \psi \right]   \ .
\end{equation}
This is easily integrated to
\begin{equation}
 S^{(2)} = {3\omega +4 \over 5\omega + 8} \chi \int d^3x \sqrt{\gamma}
           [\phi R(\gamma)
           - {\omega \over \phi} \gamma^{ij} \partial_i \phi
           \partial_j \phi ] \ ,
\end{equation}
where we ignored the irelevant homogeneous solution. Now the equation of motion
 becomes
\begin{eqnarray}
\dot \phi &=& {1 \over 2\omega +3} {1\over \sqrt{\gamma}}[ \phi
            {\delta S \over  \delta \phi} - \gamma_{ij}
              {\delta S \over  \delta \gamma_{ij}} ] \\
          &=& {H\over 2\omega +3} \phi +
          {1 \over 2\omega +3} {1\over \sqrt{\gamma}}[ \phi
            {\delta S^{(2)} \over  \delta \phi}
         - \gamma_{ij}{\delta S^{(2)} \over  \delta \gamma_{ij}} ] \ .
\end{eqnarray}
We expand $\phi$ as $\phi = \phi^{(0)} + \phi^{(2)} + \cdots$.
\begin{equation}
     \dot \phi^{(2)}
       = {H\over 2\omega +3} \phi^{(2)}
          +{3\omega +4 \over 5\omega + 8}{ \chi \over 2\omega +3}
             [ {1\over 2}\tilde{\phi} R
      -{1\over 2} {\omega \over \tilde{\phi}}\partial_k \tilde{\phi}
     \partial^k \tilde{\phi} + (2\omega +2) \Delta \tilde{\phi} ]   \ .
\end{equation}
Up to second order, the  solution is given by
\begin{equation}
  \phi = \chi^{2\over 3\omega +4} \tilde \phi
        + {(3\omega +4)^2 \over (5\omega +8)(2\omega + 3)(2\omega +4)}
             \chi^{2\omega +6 \over 3\omega +4} F(h,\tilde \phi) \ ,
\end{equation}
where
\begin{equation}
    F(h,\tilde \phi ) = {1\over 2} \tilde \phi R(h)
       -{1\over 2} {\omega \over \tilde \phi} h^{ij}
        \partial_i \phi \partial_j \phi
            + (2\omega +2) h^{ij} \tilde \phi_{;ij}  \ .
\end{equation}
Similarly, the metric is solved as
\begin{equation}
\gamma_{ij} = \chi^{4\omega +4 \over 3\omega +4} h_{ij}
    + {(3\omega +4)^2 \over (5\omega +8)(2\omega +4)}
               \chi^2 P(h,\tilde \phi)  \ ,
\end{equation}
where
\begin{eqnarray*}
   P(h,\tilde \phi)
     &=& {\omega +1 \over 2\omega +3} h_{ij} R(h) - 2 R_{ij}(h)
    - {\omega +1 \over 2\omega +3 }h_{ij}{\omega \over \tilde \phi^2}
          \partial_k \tilde \phi \partial_l \tilde \phi h^{kl}  \\
    & & \qquad + {2\omega \over \tilde \phi^2}
    \partial_i \tilde \phi \partial_j \tilde \phi
     -{2\omega +2 \over 2\omega +3}
        h_{ij}{1\over \tilde \phi}  h^{kl} \tilde\phi_{;kl}
            + {2\over \tilde \phi} \tilde \phi_{;ij} \ .
\end{eqnarray*}
The inhomogeneities of the space-time grow as
$\chi^{2\omega +4 /2\omega +3}$
 in this case. It is interesting to see if this tendency will
 continue or not.
 Therefore, we proceed to higher order calculations.
 Higher order generating functional is obtained by
\begin{equation}
  {2\omega +2 \over 2\omega +3} H \gamma_{ij}
   {\delta S^{(2n)} \over \delta \gamma_{ij} }
   + {1\over 2\omega +3} H \phi {\delta S^{(2n)} \over \delta \phi}
   + {\delta S^{(2n)} \over \delta \chi} + R^{(2n)}   = 0  \ ,
\end{equation}
where
\begin{eqnarray}
  R^{(2n)}  &=&  \sum_{k=1}^{n-1} {1\over \phi \sqrt{\gamma}}
    {\delta S^{(2k)} \over \delta \gamma_{ij} }
    {\delta S^{(2n-2k)} \over \delta \gamma_{kl} }
    \left[ \gamma_{ik} \gamma_{jl} - {\omega +1 \over 2\omega +1}
      \gamma_{ij} \gamma_{kl} \right] \\
   & &  + \sum_{k=1}^{n-1} {1\over 2(2\omega +3)} {\phi\over \sqrt{\gamma}}
    {\delta S^{(2k)} \over \delta \phi }
    {\delta S^{(2n-2k)} \over \delta \phi }
     -\sum_{k=1}^{n-1} {1\over 2\omega +3}
     {1\over \sqrt{\gamma}} \gamma_{ij}
    {\delta S^{(2k)} \over \delta \gamma_{ij} }
    {\delta S^{(2n-2k)} \over \delta \phi }   \ . \nonumber
\end{eqnarray}
Using the conformal transformation method, we obtain the recursion relation
\begin{equation}
  {\delta S^{(2n)} \over \delta \chi} + R^{(2n)} =0  \ .
\end{equation}
This leads to
\begin{eqnarray}
   S^{(2n)} &=& - \int d^3 x \int_0^1 ds \chi R^{(2n)}
          \left[s\chi \ , \psi (x) \ , f_{ij} (x) \right]  \nonumber \\
       &=& - {3\omega +4 \over (2n+3)\omega +4n+4} \chi
        \int d^3 x R^{(2n)} \left[ \chi \ , \phi (x) , \gamma_{ij} (x) \right]
\ .
\end{eqnarray}
{}From this expression, we can guess the following formal expansion
\begin{equation}
  \gamma_{ij}  =  \sum_n C_n \chi^{{4\omega +4 \over 3\omega +4}
                  + {2\omega +4 \over 3\omega +4}n} \ ,
\end{equation}
where $C_n$ can be ,in principle, determined perturbatively.
This formal expansion indicates the growing nature of the inhomogeneities of
 the space-time. It should be noted that this expression (46) coincides with
 that of the general relativity in the limit of $\omega \rightarrow \infty $.

\section{Cosmological Constant}

It is possible to show that the inhomogeneities grow or decay, as time
increases,
 depending on the equation of the state for the perfect fluid matter.
 As we investigated the dust matter which satisfies the strong energy
condition,
 here, we will study the cosmological constant model as a typical one which
does not
 satisfy the strong energy condition.
 The action is given by
\begin{equation}
I= \int d^4x \sqrt{-g} [ \phi ^{(4)}R - {\omega \over \phi} g^{\mu\nu}
     \partial_\mu \phi  \partial_\nu \phi
      - 2\Lambda ]  \ .
\end{equation}
As is explained in the previous section, we can obtain the Hamilton-Jacobi
 equation
\begin{eqnarray}
   & &{1\over \phi \sqrt{\gamma} } {\delta S \over \delta \gamma_{ij}}
           {\delta S \over \delta \gamma_{kl}}
        [\gamma_{ik} \gamma_{jl} - {\omega +1 \over 2\omega +3} \gamma_{ij}
\gamma_{kl} ]
    + {1\over 2(2\omega +3) } {\phi \over \sqrt{\gamma}}
                     ({\delta S \over \delta \phi})^2 \\
       & & \qquad  - {1\over 2\omega +3} {1\over \sqrt{\gamma} }
             \gamma_{ij} {\delta S \over \delta \gamma_{ij} }
               {\delta S \over \delta \phi}
            - \sqrt{\gamma} \phi R
           + \omega {\sqrt{\gamma} \over \phi} \gamma^{ij} \partial_i \phi
\partial_j \phi
           + 2\sqrt{\gamma} \Delta \phi + 2\Lambda \sqrt{\gamma} =0  \ .
\nonumber
\end{eqnarray}
It is our task to solve the above Hamilton-Jacobi equation using the
long-wavelength
 approximation.
The lowest equation becomes
\begin{eqnarray}
  & &{1\over \phi \sqrt{\gamma} } {\delta S^{(0)} \over \delta \gamma_{ij}}
           {\delta S^{(0)} \over \delta \gamma_{kl}}
      \left[\gamma_{ik} \gamma_{jl} - {\omega +1 \over 2\omega +3} \gamma_{ij}
\gamma_{kl} \right]
    + {1\over 2(2\omega +3) } {\phi \over \sqrt{\gamma}}
                     ({\delta S^{(0)} \over \delta \phi})^2  \nonumber \\
       & & \qquad  - {1\over 2\omega +3} {1\over \sqrt{\gamma} }
             \gamma_{ij} {\delta S^{(0)} \over \delta \gamma_{ij} }
               {\delta S^{(0)} \over \delta \phi}
            + 2\Lambda \sqrt{\gamma} =0  \ .
\end{eqnarray}
Here, we seek the quasi-isotropic solution again.
For this purpose, we take the ansatz
\begin{equation}
  S^{(0)} = - 2\sqrt{2\omega +3 \over 3\omega} \int d^3x \sqrt{\gamma} H(\phi)
\ .
\end{equation}
Substituting eq.(50) into eq.(49), we get
\begin{equation}
 {H^2 \over \phi^2} = {2\over 3\omega} ({\partial H \over \partial \phi})^2
           - {2\over \omega}{H\over \phi}{\partial H \over \partial \phi}
                      + {2\Lambda \over \phi} \ .
\end{equation}
Then, we obtain
\begin{equation}
  H= \sqrt{12\omega \Lambda \over 6\omega + 5} \phi^{1\over 2} \ .
\end{equation}
Using the equations of motion, we obtain the lowest solution
\begin{eqnarray}
  \phi^{(0)} &=&  {4\Lambda \over (2\omega+3)(6\omega+5)} t^2  \ ,  \\
  \gamma_{ij}^{(0)} &=& t^{2\omega +1} h_{ij}(x)  \ .
\end{eqnarray}
If $\omega > 1/2$, the space-time shows power-law expansion which is closely
 related to the extended inflationary scenario.
The next order equation is
\begin{eqnarray}
 & &  (2\omega+1) \sqrt{ 4\Lambda \over (2\omega +3)(6\omega +5) }
     \phi^{-{1\over 2}} \gamma_{ij} {\delta S^{(2)} \over \delta \gamma_{ij}}
      +2\sqrt{ 4\Lambda \over (2\omega +3)(6\omega +5) }
       \phi^{1\over 2}  {\delta S^{(2)} \over \delta \phi} \nonumber \\
    & & \qquad -\sqrt{\gamma} \phi R + \sqrt{\gamma}{\omega \over \phi}
      \gamma^{ij} \partial_i \phi \partial_j \phi + 2\sqrt{\gamma}
                 \Delta \phi =0 \ .
\end{eqnarray}
The direct integration yields
\begin{equation}
  S^{(2)} = {2\over (2\omega +7)A}  \int d^3x \sqrt{\gamma}
        [ \phi^{3\over 2} R  - {(\omega -1) \over \phi^{1\over 2}}
   \gamma^{ij} \partial_i \phi \partial_j \phi ]  \ ,
\end{equation}
where
\begin{equation}
  A= \sqrt{ 4\Lambda \over (2\omega +3)(6\omega +5)} \ .
\end{equation}
{}From the generating functional, the next order correction is calculated as
\begin{eqnarray}
   \phi^{(2)} &=& - {4\Lambda \over (\omega -1)(2\omega+7)
            (2\omega +3)^2 (6\omega +5) } t^{3-2\omega} R(h)  \ , \\
    \gamma_{ij}^{(2)} &=&  t^2
     [ - {2\omega +1 \over 2(\omega -1)(2\omega +3)(2\omega +7)} h_{ij} R(h)
       + {4 \over (2\omega -1)(2\omega +7)} R_{ij}(h) ] \ .
\end{eqnarray}
In contrast to the dust case, the inhomogeneities will decay if $\omega >1/2$.
This case corresponds to power law inflation.
Recursive calculation gives the higher order solutions.
It is not difficult to guess the expansion form for further correction:
\begin{equation}
  \phi = \sum t^{2+n(1-2\omega) } \bar{C_n}  \ .
\end{equation}
Therefore, the inhomogeneties of the gravitational constant will decay in the
power
 law inflation case.

\section{Discussions}
It is Dirac who first takes smallness of the following number seriously:
$$ { G m_p m_e \over e^2} \sim 10^{-40} \ , $$
where $G, e, m_p $ and $m_e$ are the gravitational constant, the charge of the
electron, the proton mass and the electron mass.
Dirac' idea is simple; The gravitational constant varies with time like as
$$ G \propto {1 \over t} \ . $$
The smallness of the number merely indicates our Universe is old!!

As a concrete model that realizes Dirac's idea, it is known that Brans-Dicke
 theory is most permissible in the sense that it appears as a low energy theory
 of superstring theory naturally and it passed the classical tests such as the
 equivalence principle.
In the isotropic and homogeneous flat Universe, the time variation of the
gravitational constant becomes
$$  {\dot G \over G} = - {H(t) \over \omega +1 }  \propto {1\over t} \ , $$
where $H$ is the Hubble parameter.

 Now we shall discuss the evolution of the gravitational constant in the
inhomogeneous
 case. First, let us consider the dust model and define the local Hubble,
\begin{eqnarray}
 \tilde H &=& {1\over 6} \gamma^{ij} \dot \gamma_{ij}  \nonumber \\
       &=& {2\omega +2 \over 3\omega +4}{1\over \chi}
 \{ 1-{(\omega +3)(3\omega +4)^2 \over 4 (5\omega +8)(2\omega +3)(\omega +1)}
 \chi^{2\omega +4\over 3\omega +4}  U  \}  \ ,
\end{eqnarray}
where
\begin{equation}
    U=  R(h) -{\omega \over \tilde{\phi}^2} h^{kl}
       \partial_k \tilde{\phi} \partial_l \tilde{\phi}
    +{2\omega \over \omega +3} {1\over \tilde{\phi}} h^{kl}
             \tilde{\phi}_{;kl} \ .
\end{equation}
Then, within the accuracy of the present approximation, the rate of variation
of gravitational constant
 is given by

\begin{equation}
  {\dot G \over G } = - {\tilde H \over \omega +1} [1 + {(3\omega +4)^2 \over
6(5\omega +8)(\omega +1)}
                          \chi^{2\omega +2 \over 3\omega +4} Q(h,\tilde \phi)]
\ ,
\end{equation}
where
\begin{equation}
  Q(h,\tilde \phi)  = R(h) - {\omega \over \tilde\phi^2} h^{ij}\partial_i
\tilde \phi
                    \partial_j \tilde \phi + {3\omega +2 \over  \tilde \phi}
                        h^{ij} \tilde \phi_{;ij}  \ .
\end{equation}
Due to the local curvature and the inhomogeneities of the Brans-Dicke field,
 the evolution of the effective coupling constant at that point is altered as
 in eq.(62). This effect becomes dominant in the late stage. For example,
 in the case $\tilde\phi =0$, the positive curvature enhanced the decreasing
 effect, and the negative curvature surpressed the decreasing effect.
 In general, the inhomogeneities of the gravitational constant itself
 gives influence to the evolution rate of $G$.
 Thus the inhomogeneities becomes significant.
  Let us look at the evolution of the density of the dust
\begin{eqnarray}
\rho &\sim &  \gamma^{-{1\over 2}} \nonumber \\
     &=& {1\over \sqrt{h}}
  \chi^{-{6\omega +6 \over 3\omega +4}} \{
  1+{(\omega +3)(3\omega +4)^2 \over 4 (5\omega +8)(2\omega +3)(\omega +2)}
 \chi^{2\omega +4\over 3\omega +4} U  \}             \ .
\end{eqnarray}
This shows that high density region and the small gravitational constant region
 coincides. This may give some interesting consequences to astrophysics.

In the case of the cosmological constant, the time evolution of the
gravitational
 constant is given by
\begin{equation}
 {\dot G \over G } = -{4\over 2\omega +1} \tilde{H}
   \left[ 1+ {(4\omega -1)(4\omega +3) \over 6(\omega -1)(2\omega +1)
         (2\omega +3)(2\omega +7) } t^{1-2\omega} R(h) \right]  \ ,
\end{equation}
where
\begin{equation}
  \tilde{H} = {2\omega +1 \over 2t } \left[
   1- { 2\omega (\omega +2) \over
        3(\omega -1)(2\omega +1)(2\omega +3)(2\omega +7)} t^{1-2\omega} R(h)
\right] \ .
\end{equation}
In the case of the power law inflation, the homogenization of the
 time variation rate of gravitational constant occurs.

\section{Conclusion}

  As a fundamentally important theory, we have studied the Brans-Dicke theory
using
 the long-wavelength approximation. First of all, we presented the
Hamilton-Jacobi
 equation for the Brans-Dicke theory which has different structure from the
 that of general relativity. The Hamilton-Jacobi equation thus obtained is
considered
 as the semi-classical equation correspondent to the Wheeler-DeWitt equation
 for the Brans-Dicke theory. Applying the method of Salopek and co-orthers,
 we have obtained the approximate solutions up to the first order of curvature
 correction in the case of dust fluid and the cosmological constant.
 As the Brans-Dicke scalar can be regarded as the effective gravitational
 constant, the non-linear evolution of the inhomogeneities of the
 Brans-Dicke field is interesting. In the above two cases, we investigated
 this problem. From the results, we can conclude that the inhomogenieties of
the
 gravitational constant will decay in the case of inflatinary matter and
 grow in the case of ordinary matter which satisfies the dominant energy
 condition.
  We have also calculated the approximate solutions using the direct method
 of Comer et.al. in the Appendix.

\vskip 1cm
\noindent
{\large \sl Acknowledgments}

 The authors would like to thank  N. Deruelle for giving us a seminar
 about long-wavelength approximation in general relativity.
   This work was supported in part by Monbusho Grant-in-Aid for Scientific
Research
 No.06740222.

\vskip 1cm
\noindent
{\large \bf APPENDIX:
                Anti-Newtonian Formalism}

 The equations of motion for the Brans-Dicke theory is
\begin{eqnarray}
G_{\mu\nu} &=& T_{M\mu\nu} +{1\over \phi} ( \phi_{;\mu\nu} - g_{\mu\nu} \Box
\phi )
            + {\omega \over \phi^2}(\partial_\mu \phi \partial_\nu \phi
        - {1\over 2} g_{\mu\nu} \partial^\alpha \phi \partial_\alpha \phi ) \\
    \Box \phi &=&  {8\pi \over 2\omega +3} T_M \ ,
\end{eqnarray}
where $T_{M\mu\nu}$ is the energy momentum tensor of the matter, hereafter we
 will consider the irrotational dust fluid.
In the synchronous gauge, we can write the above equations in the following
way:
\begin{eqnarray}
    {\partial \over \partial t } K + K^i_k K^k_i
        &=&   {1\over \phi} ( -{3\over 2} \ddot{\phi}
         -{1\over 2} K \dot \phi + {1\over 2} \Delta \phi )
         - {\omega \over \phi^2} {\dot{\phi}}^2 - {4\pi \over \phi} \rho     \\
     -K^k_{i|k} + K_{|i} &=&  - {1\over \phi} \dot\phi_{,i}
             + {1\over \phi} K^k_i \phi_{,k}
          - {\omega \over \phi^2} \dot\phi \partial_i \phi   \\
   KK^i_k + {\partial \over \partial t} K^i_k + ^{(3)}R^i_k
       &=&  {1\over \phi} \left[ \phi^{;i}_{;k} -K^i_k \dot\phi
  + {1\over 2} \delta^i_k ( -\ddot\phi -K \dot\phi +\Delta\phi ) \right]
\nonumber \\
     & & \qquad + {\omega \over \phi^2}\partial^i \phi \partial_k \phi
      + {4\pi \over \phi} \rho \delta^i_k    \\
   - \ddot{\phi} -K\dot\phi + \Delta \phi    &=& - {8\pi \over 2\omega +3} \rho
\end{eqnarray}
where $\rho$ is the energy density of the dust fluid.

Define
\begin{equation}
   K^i_j  = \Sigma^i_j + {1\over 3} \delta^i_j K
\end{equation}
We can write the above equations in the following way:
\begin{eqnarray}
  - \Sigma^k_{i|k} + {2\over 3} K_{|i} &=&
       - {1\over \phi}\dot \phi_{,i} + {1\over \phi} K^k_i \phi_{,k}
           - {\omega \over \phi^2} \dot\phi \partial_i \phi   \\
    {\partial \over \partial t} K + {1\over 2} K^2
    +{3\over 4} \Sigma^i_k \Sigma^k_i  + {1\over 4} ^{(3)} R
     &=& -{1\over \phi} ({3\over 2} \ddot \phi + K \dot \phi )
        +{1\over \phi} \Delta \phi  \nonumber \\
     & & \qquad -{3\over 4} {\omega \over \phi^2}\dot\phi^2
     + {\omega \over 4} {1\over \phi^2} \partial^k \phi \partial_k \phi \\
  {\partial \over \partial t} K + \Sigma^i_k \Sigma^k_i
  +{1\over 3} K^2 &=& - {\omega \over \phi^2} \dot\phi^2
   -  (\omega +3) {\ddot\phi \over \phi } \nonumber \\
   & & \qquad - (\omega +2) K {\dot\phi \over \phi}
       +  (\omega +2) {1\over \phi} \Delta \phi  \\
  {\partial \over \partial t} \Sigma^i_k  + K \Sigma^i_k
   - \Sigma^i_k {\dot\phi \over \phi} &=&
     R^i_k -{1\over 3} \delta^i_k R +
     {1\over \phi}(\phi^{;i}_{;k} -{1\over 3} \delta^i_k \Delta \phi )
\nonumber \\
  & & \qquad  + {\omega \over \phi^2} ( \partial^i \phi \partial_k \phi
   -{1\over 3} \delta^i_k \partial^m \phi \partial_m \phi )
\end{eqnarray}
We now consider these equations order by order in the gradient expansion.
At zeroth order, we neglect the terms which have twice derivative,
 i.e., $^{(3)} R , \Delta \phi$ etc.. Futhermore, we impose quasi-isotropic
nature
 $\Sigma^i_j =0 $.
 In this order, we obtain the solutions
\begin{eqnarray}
   \gamma_{ij} &=& t^{4\omega +4 \over 3\omega +4} h_{ij}(x) \ , \\
   \phi &=& \tilde{ \phi} (x) t^{2\over 3\omega +4}   \ , \\
   \rho &=&  {\tilde{ \phi} (x) \over 8\pi}
         {4\omega + 6 \over 3\omega +4} t^{-{6\omega +6 \over 3\omega +4}} \ .
\end{eqnarray}
Here, $h_{ij}$ and $\tilde{\phi}$ depend on the spatial coordinates.
Substituting (79) $\sim$ (81) into eqs.(75)$\sim$ (78) and keping the next
oeder terms,
 we obtain
\begin{eqnarray}
  \Sigma^i_j &=& {3\omega +4 \over 5\omega +8} t^{-{\omega \over 3\omega +4}}
   [  -( R^i_j - {1\over 3} \delta^i_j R )
     +{1\over \phi} ( \phi^{;i}_{;j}
      - {1\over 3} \delta^i_j  \Delta \phi ) \nonumber \\
 & & \qquad  +{\omega \over \phi^2} ( \partial^i \phi \partial_j \phi
    - {1\over 3} \delta^i_j  \partial^k \phi \partial_k \phi ) ]
\end{eqnarray}

\begin{equation}
  \phi^{(2)} =  {(3\omega +4)^2 \over (5\omega +8)(2\omega +4)(2\omega +3)}
   t^{2\omega +6 \over 3\omega +4} \left[ {1\over 2} \phi R
  + 2 (\omega +1) \Delta \phi
  -{\omega \over 2 \phi} \partial^k \phi \partial_k \phi \right]
\end{equation}

\begin{eqnarray}
  K^{i(2)}_j &=&  {3\omega +4 \over 5\omega +8} t^{-{\omega \over 3\omega +4}}
   [  -R^i_j + {\omega +1 \over 2(2\omega +3)} \delta^i_j R
     +{1\over \phi} \phi^{;i}_{;j}
      -{\omega +1 \over 2\omega +3} \delta^i_j {1\over \phi} \Delta \phi
\nonumber \\
 & &\qquad \qquad +{\omega \over \phi^2} \partial^i \phi \partial_j \phi
    -{\omega +1 \over 2(2\omega +3)} \delta^i_j
   {\omega \over \phi^2} \partial^k \phi \partial_k \phi ]
\end{eqnarray}

\begin{eqnarray}
  \gamma_{ij}^{(2)} &=&  {(3\omega +4)^2 \over (5\omega +8)(\omega +2)} t^{2}
   [  -R^i_j + {\omega +1 \over 2(2\omega +3)} \delta^i_j R
     +{1\over \phi} \phi^{;i}_{;j}
      -{\omega +1 \over 2\omega +3} \delta^i_j {1\over \phi} \Delta \phi
\nonumber \\
 & &\qquad \qquad +{\omega \over \phi^2} \partial^i \phi \partial_j \phi
    -{\omega +1 \over 2(2\omega +3)} \delta^i_j
   {\omega \over \phi^2} \partial^k \phi \partial_k \phi ]
\end{eqnarray}
The above results completely agree with the Hamilton-Jacobi calculation.


\vskip 0.5cm
\begin{thebibliography}{99}
%

\bibitem{damour}
T. Damour and A. M. Polyakov, \sl Nucl. Phys. \bf B423 \rm (1994) 532.
\bibitem{BD}
C. Brans and R. H. Dicke,, \sl Phys. Rev.  \bf 124 \rm (1961) 925.
\bibitem{linde}
J. Garcia-Bellido, A. Linde and D. Linde, \sl Phys. Rev. \bf D50 \rm (1994)
730.
\bibitem{SB}
D. S. Salopek and J.R.Bond, \sl Phys. Rev. \bf D42 \rm (1990) 3936.
\bibitem{LK}
E. M. Lifschitz and I. M. Khalatnikov, \sl Adv. Phys.  \bf 12 \rm (1963) 185.
\bibitem{tomita}
K. Tomita, \sl Prog. Theor. Phys. \bf 48 \rm (1972) 1503;
K. Tomita, \sl Prog. Theor. Phys. \bf 54 \rm (1975) 730.
\bibitem{PSS}
J. Parry, D. S. Salopek and J. M. Stewart, \sl Phys. Rev.  \bf D49 \rm (1994)
2872.
\bibitem{CPSS}
K. M. Croudace, J. Parry, D. S. Salopek and J. M. Stewart, \sl Ap. J.  \bf 423
\rm (1994) 22.
\bibitem{comer}
G. L. Comer, N. Deruelle, D. Langlois and J. Parry, \sl Phys. Rev. \bf D49 \rm
(1994) 2759.
\bibitem{N}
Y.Nambu and A.Taruya, Application of Gradient Expansion to Inflationary
Universe,
 preprint (1994);
N.Deruelle and D.S.Goldwirth, Conditions for inflation in an initially
inhomogeneous
 universe, preprint (1994);
 T. Chiba, Applying Gradient Expansion to a Perfect Fluid and Higher
Dimensions,
preprint (1995).

%
\end{thebibliography}
\end{document}